\documentclass{ws-procs9x6}

\begin{document}
\title{Twisted Space-Time Symmetry, Non-Commutativity and Particle Dynamics}

\author{J. LUKIERSKI and M. WORONOWICZ}

\address{Institute of Theoretical Physics \\
50-205 Wroc\l aw, \\
pl. Maxa Borna 9, Poland\\
E-mail: lukier,woronow@ift.uni.wroc.pl}

\maketitle

\abstracts{We describe the twisted space-time symmetries which
imply the quantum Poincar\'{e} covariance of noncommutative Minkowski spaces, with
constant, Lie algebraic and quadratic commutators. Further we present the relativistic and nonrelativistic particle models 
invariant respectively under twisted relativistic and twisted Galilean 
symmetries.}

\section{Introduction}
Since the work of Doplicher et all. (see e.g.\cite{luwo1,luwo2}) there is a strong indication that due
to quantum gravity effects the space-time coordinates are becoming noncommutative.
In general case one can write\footnote{Formula (\ref{luwoeq1}) is not the most general one. One
can assume that the rhs of (\ref{luwoeq1}) depends also on momenta (or derivative operators) as well as
on other operators, e.g. spin variables. In this note we shall not consider such extensions
of (\ref{luwoeq1}). The expansion (\ref{luwoeq1}) is only up to quadratic term because higher orders do not have classical
 limit $\kappa \to \infty$.}
 \begin{eqnarray}\label{luwoeq1}
 [ \widehat{x}_{\mu},\widehat{x}_{\nu}] & = &
 \frac{i}{\kappa^2} \,  \theta_{\mu\nu} (\kappa \widehat{x}_{\rho})
 \cr
 & = &
 \frac{i}{\kappa^2} \,  \theta_{\mu\nu}^{(0)}
 +
 \frac{i}{\kappa} \,  \theta_{\mu\nu}^{(1) \rho } \,
 \widehat{x}_\rho
 + i
   \theta_{\mu\nu}^{(2) \rho \tau}
   \widehat{x}_{\rho}\widehat{x}_{\tau}\, ,
 \end{eqnarray}
where the fundamental mass parameter $\kappa$ has been introduced in order to exhibit the mass
dimensions of respective terms and have the constant tensors $ \theta_{\mu\nu}^{(0)}$,
 $ \theta_{\mu\nu}^{(1) \rho}$, $ \theta_{\mu\nu}^{(2) \rho \tau}$ as dimensionless.
 If we link (\ref{luwoeq1}) with quantum gravity one can put $\kappa = m_{\rm pl}$
  ($m_{\rm pl}$ - Planck mass). Further we add that the relation (\ref{luwoeq1}) describes in D=10 first-quantized open string theory the
  noncommutative coordinates on D-branes providing the localizations of the ends of the strings
   \cite{sw,chs}.

   There are two important problems related with the application of formula (\ref{luwoeq1})
    to physical models:

{\bf     i)}
    In standard relativistic theory, with classical Poincar\'{e} symmetries, the first
     term on rhs of (\ref{luwoeq1}) breaks the Lorentz invariance, and further two terms break both
     Lorentz and translational invariance. One can ask how looks the deformation of classical
     Poincar\'{e} invariance which permits to consider relations (\ref{luwoeq1}) as covariant under deformed Poincar\'{e} transformations, i.e.  the same in any deformed Poincar\'{e} frame.

{\bf      ii)}
      There should be given prescriptions how to formulate the classical mechanics and
      field theory models with noncommutative space-time coordinates (\ref{luwoeq1}), covariant under the twisted Poincar\'{e} symmetries. 
      
If the time coordinate remains classical (i.e. in formula (\ref{luwoeq1}) 
$\theta_{0\mu} = 0$) both  points i) and ii) can be applied to the
nonrelativistic noncommutative theories with classical Galilean invariance 
broken by relation (\ref{luwoeq1}).

\section{Twisted Space-Time Symmetries}

We shall look for the quantum relativistic symmetries implying the covariance of noncommutative Minkowski spaces. In systematic study firstly one should consider all possible quantum relativistic symmetries (quantum Poincar\'{e} algebras) in the form of noncommutative Hopf algebras, and then derive corresponding quantum Minkowski spaces as deformed Hopf algebra modules. An example of such a construction which is already more than ten years old is the $\kappa$-deformed Minkowski space \cite{zak,mr,lrz}
	\begin{equation}\label{kappa}
		[\widehat{x}_0,\widehat{x}_i]=\frac{i}{\kappa}\widehat{x}_i\,,\qquad\qquad [\widehat{x}_i,\widehat{x}_j]=0\,,
	\end{equation}
corresponding in (\ref{luwoeq1})  to the choice $\theta^{(0)}_{\mu\nu}=\theta^{(2)\rho\tau}_{\mu\nu}=0$ and 
$\theta^{(1)\rho}_{\mu\nu}=\eta_{\nu 0}\delta_\mu^{\ \rho}-\eta_{\mu 0}\delta_{\nu}^{\ \rho}$. Using the Hopf-algebraic formulae of $\kappa$-deformed Poincar\'{e} algebra in bicrossproduct basis one can show \cite{mr} that the relations (\ref{kappa}) are covariant under the Hopf-algebraic action of $\kappa$-deformed Poincar\'{e} algebra.

It appears that the most effective way of describing the noncommutative space-times covariant under quantum relativistic symmetries is to consider twisted symmetry algebras. In such a case the classical Poincar\'{e}-Hopf algebra is modified only in the coalgebraic sector, with all the algebraic relations preserved. We change the classical Poincar\'{e} Hopf algebra $\mathcal{H}^{(0)}=(\mathcal{U}(\mathcal{P}_4), m, \Delta_0, S_0, \epsilon)$ into twisted Poincar\'{e} Hopf algebra $\mathcal{H}=(\mathcal{U}(\mathcal{P}_4), m, \Delta, S, \epsilon)$ by means of the twist factor $\mathcal{F}\in\mathcal{U}(\mathcal{P}_4)\otimes\mathcal{U}(\mathcal{P}_4)$ as follows 
($\mathcal{P}_4\ni\hat{g}=(P_\mu, M_{\mu\nu})$)
\begin{equation}
	\Delta(\hat{g})=\mathcal{F}\circ\Delta_0(\hat{g})\circ\mathcal{F}^{-1}\,,\qquad S(\hat{g})=US_0(\hat{g})U^{-1}\,,
\end{equation}
\begin{equation}
	\Delta_0(\hat{g})=\hat{g}\otimes 1+1\otimes\hat{g}\,,\qquad S_0(\hat{g})=-\hat{g}\,,\qquad \epsilon(\hat{g})=0\,,
\end{equation}
where $(a\otimes b)\circ(c\otimes d)=ac\otimes bd$. The twist $\mathcal{F}$ satisfies the cocycle and normalization conditions \cite{dr}
\begin{equation}\label{cocy}
\mathcal{F}_{12}\,\left(\Delta_0\otimes 1\right)\,\mathcal{F}=\mathcal{F}_{23}\,\left(1\otimes\Delta_0 \right)\,\mathcal{F}\,,\qquad (\epsilon\otimes 1)\mathcal{F}=(1\otimes\epsilon)\mathcal{F}=1\,,
\end{equation}
where $\mathcal{F}_{12}=\textup{f}_{(1)}\otimes \textup{f}_{(2)}\otimes 1$ etc. $(\mathcal{F}=\textup{f}_{(1)}\otimes \textup{f}_{(2)})$ and $U= \textup{f}_{(1)}S(\textup{f}_{(2)})$.

The advantage of using twisted Poincar\'{e} algebra is the explicit formula for the multiplication in twisted Hopf algebra module $\mathcal{A}$ which should satisfy the condition (see e.g. \cite{majid}, $h\in\mathcal{U}(\mathcal{P}_4),\ a,b\in\mathcal{A}$)
\begin{equation}\label{module}
	h\rhd(a\bullet b)=(h_1 a)\bullet(h_2 b)\,,
\end{equation}
where $\Delta(h)=h_1\otimes h_2$. We see from (\ref{module}) that if $h_1\neq h_2$ then $a\bullet b\neq b\bullet a$, i.e. from quantum-deformed relativistic symmetry follow necessarily the noncommutative Minkowski space as its Hopf-algebraic module.

One can show that the multiplication in $\mathcal{A}$ for twisted Hopf algebra $\mathcal{H}$ which is consistent with the relation (\ref{module}) ($h\in\mathcal{H}$) provides the formula \cite{km,bloh,cknt}
\begin{equation}\label{modtw}
	a\bullet b=(\overline{\textup{f}}_{(1)} a)(\overline{\textup{f}}_{(2)} b)\,,\qquad \mathcal{F}^{-1}=\overline{\textup{f}}_{(1)}\otimes\overline{\textup{f}}_{(2)}\,.
\end{equation}
In the case of relativistic symmetries one can use the classical space-time representation for the Poincar\'{e} generators $P_\mu,\ M_{\mu\nu}$
\begin{equation}\label{repgen}
	P_\mu=i\partial_\mu\,,\qquad M_{\mu\nu}=i(x_\nu\partial_\mu-x_\mu\partial_\nu)\,.
\end{equation}
Subsequently in the formula (\ref{modtw}) one can assume that $a,\ b$ are classical functions on commutative Minkowski space $x_\mu$, and define $\overline{\textup{f}}_{(i)}(P_\mu,\ M_{\mu\nu})\equiv \overline{\mathfrak{f}}_{(i)}(x,\ \partial),\  i=1,2$. One gets the following star product multiplication which is a particular representation of algebraic formula (\ref{modtw}) 
\begin{equation}
	\xi(x)\star\zeta(x)=(\bar{\mathfrak{f}}_{(1)}(x,\partial)\xi(x)) (\bar{\mathfrak{f}}_{(2)}(x,\partial)\zeta(x))\,.	
\end{equation}

The important application of twisted Poincar\'{e} algebras to the covariant description of noncommutative Minkowski spaces, namely describing the quantum covariance of (\ref{luwoeq1}) for the case $\theta_{\mu\nu}=\theta^{(0)}_{\mu\nu}$ is quite recent\footnote{The twisted Poincar\'{e} symmetries corresponding to $\theta_{\mu\nu}=\theta^{(0)}_{\mu\nu}$ were earlier discussed in \cite{asch96,asch99,oeckl,js}, but the full consequences of the twisted description were realized in 2004 (see e.g. \cite{cknt,jwess,kt,koma}).}. The quantum symmetry which leaves invariant the simplest form of (\ref{luwoeq1})\footnote{Below, in chapter 2 and 3, we shall use explicitly the \textit{fat dot notation} for the algebra of functions on quantum Minkowski space in order to stress its Hopf algebra module origin.}
\begin{equation}\label{thetacom}
	[\widehat{x}_\mu,\widehat{x}_\nu]_\bullet=\frac{i}{\kappa^2}\theta^{(0)}_{\mu\nu}\,,
\end{equation}
(where $[a,b]_\bullet=a\bullet b-b\bullet a$) is generated by the following Abelian twist
\begin{equation}\label{twist0}
\mathcal{F}_{\theta }=\exp \,\frac{i}{2\kappa^2}(\,\theta ^{\mu \nu }_{(0)}\,P_{\mu }\wedge P_{\nu
}\,)\,.  
\end{equation}
We obtain the twisted Poincar\'{e}-Hopf structure with classical Poincar\'{e} algebra relations and modified coproducts of Lorentz generators $M_{\mu\nu}$
\begin{eqnarray}\label{cop0}
\Delta_\theta(P_\mu)&=&\Delta_0(P_\mu), \\
\Delta _{\theta }(M_{\mu \nu })&=&\mathcal{F}_{\theta }\circ \,\Delta _{0}(M_{\mu \nu
})\circ \mathcal{F}_{\theta }^{-1}\nonumber\\
&=&\Delta _{0}(M_{\mu \nu })-\frac{1}{\kappa ^{2}}%
\theta ^{\rho \sigma }_{(0)}[(\eta _{\rho \mu }P_{\nu }-\eta _{\rho \nu
}\,P_{\mu })\otimes P_{\sigma }\label{cop00}\\
&&\qquad\qquad\qquad\qquad\qquad+P_{\rho}\otimes (\eta_{\sigma \mu}P_{\nu}-\eta_{\sigma \nu}P_{\mu})]\,.\nonumber
\end{eqnarray}

One can consider however also other Abelian twists of Poincar\'{e} symmetries, depending on the Lorentz generators $M_{\mu\nu}$ (see \cite{lnrt,asch96,asch99,luwo}). It appears that only subclass of general commutator (\ref{luwoeq1}) with linear and quadratic terms can be covariantized by twisted Poincar\'{e} algebras. In the following section we shall consider the quantum Poincar\'{e} symmetries corresponding to the following two twist functions \cite{luwo}:
\begin{itemize}
\item[i)] Lie-algebraic relations for noncommutative Minkowski space
\begin{equation}\label{twist1}
\mathcal{F}_{(\alpha\beta)}= \exp \,\frac{i}{2\kappa}( \zeta^\lambda\,P_{\lambda }\wedge M_{\alpha \beta }) \,,
\end{equation}
where $\alpha,\ \beta=0,1,2,3$ are fixed and the vector $\zeta^\lambda=\theta^{\lambda\alpha\beta}_{(1)}$ has vanishing components $\zeta^\alpha,\ \zeta^\beta$.

\item[ii)] Quadratic deformations of Minkowski space
\begin{equation}\label{twist2}
    \mathcal{F}_{(\alpha\beta\gamma\delta)}=\exp\,\frac{i}{2}\zeta\, M_{\alpha\beta}\wedge M_{\gamma\delta}\,,
\end{equation}
where $\zeta=\theta_{(2)}^{\alpha\beta\delta\gamma}$ is a numerical parameter, all the four indices $\alpha,\beta,\gamma,\delta$ are fixed and different.
\end{itemize}

\section{Lie-algebraic and Quadratic Quantum-Covariant Noncommutative Minkowski Spaces}

In this Section we shall report on results presented in \cite{luwo}, which we supplement by the proof of quantum translational invariance. 

In the formalism of quantum-deformed Hopf-algebraic symmetries the quantum-covariant noncommutative Minkowski space can be introduced in two ways:
\begin{itemize}
	\item[i)] as the translation sector of quantum Poincar\'{e} group\,,
	\item[ii)] as the quantum representation space (a Hopf algebra module) for quantum Poincar\'{e} algebra with the action of the deformed symmetry generators satisfying suitably deformed Leibnitz rule (\ref{module}).
\end{itemize}

In the case of constant tensor $\theta_{\mu\nu}=\theta_{\mu\nu}^{(0)}$ the quantum Poincar\'{e} group algebra dual to the coproducts (\ref{cop0}), (\ref{cop00}) is known\cite{oeckl,koma,luwo}, and the quantum translations do not satisfy the relation (\ref{thetacom}). It appears that the relation (\ref{thetacom}) as describing quantum-covariant noncommutative Minkowski space can be obtained only as the Hopf algebra module. To the contrary, in the case of twisted relativistic symmetries generated by the twist factors (\ref{twist1}, \ref{twist2}) it can be shown that both definitions i) and ii) coincide \cite{luwo}.

\begin{itemize}
\item[i)] Lie-algebraic noncommutative Minkowski space.

The commutator algebra following from (\ref{twist1}) and the formula (\ref{modtw}) has the form \cite{luwo}
\begin{equation}\label{lie}
\left[ \widehat{x}_{\mu },\widehat{x}_{\nu }\right] _{\bullet}=C_{\ \mu \nu }^{\rho }\widehat{x}_{\rho }\,,
\end{equation}
where 
\begin{equation}\label{const}
	C_{\ \mu \nu }^{\rho }=\frac{i}{\kappa}\zeta_\mu( \eta _{\beta \nu }\delta
_{\ \alpha }^{\rho }-\eta _{\alpha \nu }\delta _{\ \beta }^{\rho })+\frac{i}{\kappa}\zeta_\nu( \eta _{\alpha \mu }\delta _{\ \beta }^{\rho }-\eta _{\beta \mu }\delta _{\ \alpha }^{\rho
}) \,.
\end{equation}
The relations (\ref{lie}) can be written in more transparent way as follows ($\alpha,\ \beta$ are fixed by the choice of twist function)
\begin{equation}\label{komlie}
\left[\widehat{x}_\alpha,\widehat{x}_\lambda\right]_\bullet=\frac{i}{\kappa}\zeta_\lambda \eta_{\alpha\alpha}\widehat{x}_\beta\,,\qquad\qquad
\left[\widehat{x}_\beta,\widehat{x}_\lambda\right]_\bullet=-\frac{i}{\kappa}\zeta_\lambda \eta_{\beta\beta}\widehat{x}_\alpha\,,
\end{equation}
where $\zeta_\alpha=\zeta_\beta=0$.

The quantum Lorentz covariance of (\ref{lie}) under the Hopf action of the Lorentz generators $M_{\mu\nu}$ has been shown in \cite{luwo}. We shall show the quantum translational invariance of (\ref{lie}) using the differential realization (\ref{repgen}). The fourmomentum coproduct generated by twist (\ref{twist1}) has the form \cite{luwo}
\begin{equation}
		\Delta(P_\mu)=\Delta_0(P_\mu)+\frac{1}{2\kappa}\xi^\lambda P_\lambda\wedge(\eta_{\alpha\mu} P_\beta-\eta_{\beta\mu} P_\alpha)+\mathcal{O}(P^3)\,.
\end{equation}
Putting in (\ref{module}) $h\equiv P_\mu$, $a\equiv x_\rho$, $b\equiv x_\sigma$ and using (\ref{const}) we obtain
\begin{eqnarray}
	P_{\mu}\rhd\left(x_\rho\bullet x_\sigma\right)	&=&ix_{\{\rho}\eta_{\sigma\}\mu}+\eta_{\alpha\mu}\xi_{[\sigma}\eta_{\rho]\beta}-\eta_{\beta\mu}\xi_{[\sigma}\eta_{\rho]\alpha}\,,\\
&=&ix_{\{\rho}\eta_{\sigma\}\mu}+\frac{1}{2}P_{\mu}\rhd C_{\ \rho\sigma}^{\lambda}x_\lambda\,.\nonumber
\end{eqnarray}
Finally we get
\begin{equation}
	P_\mu\rhd\left[x_\rho,x_\sigma\right]_\bullet=P_\mu\rhd C_{\ \rho\sigma}^{\lambda}x_\lambda\,,
\end{equation} 
i.e. the relation (\ref{lie}) is covariant.

\item[ii)] Quadratic noncommutativity of Minkowski space coordinates.

After using the formula (\ref{modtw}) with inserted twist (\ref{twist2}) one gets the following commutation relations of space-time coordinates ($\{a,b\}_\bullet=a\bullet b+b\bullet a$)
\begin{eqnarray}
[\widehat{x}_\mu,\widehat{x}_\nu]_\bullet&=&i\sinh\frac{\zeta}{2}\cosh\frac{\zeta}{2}
(\eta_{\alpha[\mu}\eta_{\gamma\nu]}\{\widehat{x}_\beta,\widehat{x}_\delta\}_\bullet
-\eta_{\alpha[\mu}\eta_{\delta\nu]}\{\widehat{x}_\beta,\widehat{x}_\gamma\}_\bullet\label{quadcom}\\
&&\qquad\qquad\qquad\qquad-
\eta_{\beta[\mu}\eta_{\gamma\nu]}\{\widehat{x}_\alpha,\widehat{x}_\delta\}_\bullet
+\eta_{\beta[\mu}\eta_{\delta\nu]}\{\widehat{x}_\alpha,\widehat{x}_\gamma\}_\bullet)\nonumber\\&-&\sinh^2\frac{\zeta}{2}(\sum_{\genfrac{}{}{0pt}{}{k=\alpha,\beta}{l=\gamma,\delta}}^{}\delta^k_{\ [\mu}\delta^l_{\ \nu]}[\widehat{x}_k,\widehat{x}_l]_\bullet)\,,\nonumber
\end{eqnarray}
or in more explicit form ($k=\alpha,\beta$ and $l=\gamma,\delta$)
\begin{eqnarray}\label{komquad}
&&[\widehat{x}_k,\widehat{x}_l]_\bullet=i\tanh\frac{\zeta}{2}
(\eta_{\alpha k}\eta_{\gamma l}\{\widehat{x}_\beta,\widehat{x}_\delta\}_\bullet
-\eta_{\alpha k}\eta_{\delta l}\{\widehat{x}_\beta,\widehat{x}_\gamma\}_\bullet\label{star1q}\\
&&\qquad\qquad\qquad\qquad-
\eta_{\beta k}\eta_{\gamma l}\{\widehat{x}_\alpha,\widehat{x}_\delta\}_\bullet
+\eta_{\beta k}\eta_{\delta l}\{\widehat{x}_\alpha,\widehat{x}_\gamma\}_\bullet)\,,\nonumber
\end{eqnarray}
and $[\widehat{x}_\alpha, \widehat{x}_\beta]_\bullet=[\widehat{x}_\gamma,\widehat{x}_\delta]_\bullet=0$.

We conjecture that the relations (\ref{quadcom}) are covariant under the action of quantum Poincar\'{e} symmetries, generated by twist (\ref{twist2}).

The linear and quadratic relations (\ref{komlie}) and (\ref{komquad}) provide special choices of the constant parameters $\theta^{(1)\rho}_{\mu\nu}$, $\theta^{(2)\rho\tau}_{\mu\nu}$ for which the quantum covariance group was found in \cite{luwo}.
\end{itemize}

\section{Particle Dynamics Invariant Under Twisted Relativistic and Galilean Symmetries}

The discussion of the noncommutative  dynamical theories one begins naturally with the consideration of classical mechanics models. We shall restrict our considerations here to the case $\theta_{\mu\nu}=\theta^{(0)}_{\mu\nu}$, i.e. the noncommutative space-time described by (\ref{thetacom}). One can introduce the Lagrangian models describing free point particles moving in noncommutative space-time in the following two ways:
\begin{itemize}
	\item[i)] If $\theta_{\mu 0}=0$, i.e. we have the relations
	\begin{eqnarray}
	&&[\widehat{x}_i,\widehat{x}_j]=i\theta_{ij}\,,\label{dir}\\
	&&[\widehat{x}_0,\widehat{x}_i]=0\,,
	\end{eqnarray}
	we deal with classical time variable $t$, where $\hat{x}_0=ct$ and noncommutative space coordinates $\hat{x}_i$.
	In such a case one can look for the non-relativistic Lagrangian models with constraints, which provide the relation (\ref{dir}) as the quantized Dirac bracket. Such a first model was constructed in \cite{lsz1} in $D=(2+1)$ dimensions with the following Lagrangian 
	\begin{equation}\label{lag}
		\mathcal{L}=\frac{m\dot{x}^2_i}{2}-k\epsilon_{ij}\dot{x}_i \ddot{x}_j\,.
	\end{equation}	
	The higher order Lagrangian (\ref{lag}) can be expressed if first order form in six-dimensional phase space $(x_i, p_i, \tilde{p}_i)$\footnote{The momenta $p_i,\tilde{p}_i$ are described by the following formulae
		\begin{equation}
		p_i=\frac{\partial\mathcal{L}}{\partial \dot{x}_i}-\frac{d}{dt}\frac{\partial\mathcal{L}}{\partial\ddot{x}_i}\,,\qquad \tilde{p}_i=\frac{\partial\mathcal{L}}{\partial\ddot{x}_i}\,.\nonumber
	\end{equation}
  }
  and  after introducing the linear transformations
  \begin{eqnarray}
  	&&X_i=x_i-\frac{2}{m}\tilde{p}_i\,,\nonumber\\
  	&&P_i=p_i\,,\\
  	&&\tilde{P}_i=\epsilon_{ij}\tilde{p}_j+\frac{k}{m}p_i\,,\nonumber
  \end{eqnarray}
  one obtains the following symplectic structure for the variables $Y_A=(X_i,P_i,\tilde{P}_i),\ (A=1...6)$
  \begin{equation}
  	\{Y_A,Y_B\}=\Omega_{AB}\,,\qquad\qquad \Omega=
  	\left(\begin{array}{ccc}
  							\frac{2k}{m^2}\epsilon & 1_2 & 0 \\
  							-1_2 & 0 &0 \\
  							0 & 0 & \frac{k}{2}\epsilon
  							\end{array}\right)\,.
  \end{equation}
  One can identify (\ref{dir}) with quantized PB for the space variables $X_i$ if we put in (\ref{dir}) $\theta_{ij}=\frac{2k}{m^2}\epsilon_{ij}$.
  
  In \cite{lsz1} the dimension $D=2+1$ was chosen because in two space dimensions one can put $\theta_{ij}=\theta\epsilon_{ij}$, i.e. the relation (\ref{dir}) does not break the classical Galilean invariance. However if $k\neq 0$ the Galilean algebra is centrally extended by second \textit{exotic} central charge \cite{levy}.\\
  
  \item[ii)] For general constant $\theta_{\mu\nu}$  one obtains the noncommutative action describing free particle motion if we introduce in the first order action for classical massive relativistic particle
  \begin{equation}\label{act1}
  	S=\int d\tau[\dot{y}_\mu p^\mu-e(p^2-m^2)]\,,
  \end{equation}
  the following change of variables (we recall that $\theta_{\mu\nu}=\theta_{\mu\nu}^{(0)}$)
  \begin{equation}\label{change}
  	y_\mu=x_\mu+\frac{1}{a}\theta_{\mu\nu}p^\nu\,.
  \end{equation}
  It is easy to check that if we introduce CCR following from (\ref{act1})
 \begin{equation}
  	[y_\mu,y_\nu]=0\,,\qquad [y_\mu, p^\nu]=i\delta_\mu^{\ \nu}\,, \qquad [p^\mu,p^\nu]=0\,,
  \end{equation}
  then the variables $x_\mu$ in (\ref{change}) satisfy the relation (\ref{thetacom}) if we put $a=2\kappa^2$. 
  Using the relation (\ref{change}) one can rewrite the action (\ref{act1}) as follows
  \begin{equation}\label{act2}
  	S=\int d\tau[\dot{x}_\mu p^\mu-e(p^2-m^2)+\frac{1}{a}\theta^{\mu\nu}\dot{p}_\mu p_\nu]\,.
  \end{equation}
  
  The variables $y_\mu,p_\mu$ in (\ref{act1}) are classical, i.e. transform under Lorentz rotations in standard way
  \begin{equation}\label{lorentz}
  	y'_\mu=\Lambda_\mu^{\ \nu}y_\nu\,,\qquad\qquad p'_\mu=p_\mu\,.
  \end{equation}
  Using (\ref{change}) and (\ref{lorentz}) one gets however
  \begin{eqnarray}\label{deflor}
  	x'_\mu&=&y'_\mu+\frac{1}{a}\theta_{\mu\nu}\Lambda^\nu_{\ \rho}p^\rho\\
  	&=&\Lambda_\mu^{\ \nu}x'_\nu+\frac{1}{a}(\Lambda_\mu^{\ \rho}\theta_{\rho\nu}+\theta_{\mu\rho}\Lambda^\rho_{\ \nu})p^\nu\nonumber\,.
  \end{eqnarray}
  Interestingly enough, the transformations (\ref{deflor}) describe exactly the twisted Lorentz transformations, generated by the coproduct (\ref{cop00}), which leave invariant the action (\ref{act2}) for the noncommutative relativistic particle.
	\end{itemize}
	
  The model (\ref{act2}) has been firstly obtained without reference to twisted Lorentz symmetries by Deriglazov \cite{der} and its non-relativistic version
  \begin{equation}\label{actnr}
  	S_{NR}=\int dt[\dot{x}_i\dot{p}_i-\frac{1}{2m}\vec{p}^{\ 2}+\frac{1}{a}\theta_{ij}\dot{p}_ip_j]\,,
  \end{equation}
  in $D=2+1$, when $\theta_{ij}=\epsilon_{ij}$, it was proposed by Duval and Horvathy \cite{dhor}. It is well-known however that the model (\ref{actnr}) can be also derived from the model (\ref{lag}). Indeed, the first order formulation of the model (\ref{lag}) in Faddeev-Jackiw approach \cite{fj} to higher order Lagrangians provides the action \cite{hp,lsz2} 
  \begin{equation}\label{lag3}
  	\mathcal{L}=p_i(\dot{x}_i-y_i)+\frac{\vec{y}^{\ 2}}{2m}+\frac{1}{a}\epsilon_{ij}\dot{p}_i p_j\,.
  \end{equation}
 The Lagrangian (\ref{lag3}) after introducing the new coordinates
  \begin{equation}
  		X_{i} = x_{i} + \frac{1}{a} \epsilon_{ij}(y_{j} - p_{j})\,,     
  \end{equation}
provides the Lagrangian (\ref{actnr}) (with $x_{i}$ replaced by $X_{i}$) and additional 
term which depends on auxiliary internal variables commuting with $(X_i, p_i)$ \cite{hp}.
 
The nonrelativistic model (\ref{actnr}) can be considered in any space dimension $d$ . 
If $d = 2$ the action (\ref{actnr}) is, similarly as (\ref{lag}), invariant under the 
transformations of exotic $(2+1)$ - dimensional Galilean group. 
If $d >2$ the invariance of the nonrelativistic model (\ref{actnr}) can be achieved by considering quantum Galilean symmetries, with twisted space rotations generated 
by the following nonrelativistic twist
\begin{equation}
\mathcal{F}^{NR}_{\theta} = \exp \frac{i}{a} \theta_{ij}P_i \wedge P_j\,.
\end{equation}
 The formulation of twisted quantum mechanics invariant under twisted 
quantum Galilei group is now under our consideration.

\section{Final Remarks}

We presented in this paper some selected aspects of the theory of noncommutative space-times, with new results on  quantum Poincar\'{e} covariance of a class of linearly and quadratically deformed Minkowski spaces. We also considered the non-relativistic and relativistic particle models on noncommutative space-time with numerical value of the noncommutativity function $\theta_{\mu\nu}=\theta_{\mu\nu}^{(0)}$ and have pointed out their twisted quantum covariance. We see that the role of quantum deformations is to introduce 
in place of broken classical symmetries a modified transformations which
imply the quantum covariance. Such a possibility selects only particular 
class of tensors $\theta^{(1)\rho}_{\mu\nu}$ and $\theta^{(2)\rho\tau}_{\mu\nu}$ in formula (\ref{luwoeq1}).

Most of the applications of the noncommutative space-times in the literature assume the choice $\theta_{\mu\nu}=\theta^{(0)}_{\mu\nu}$ (see (\ref{thetacom})). In this talk we presented also the results for linear 
($\theta^{(1)\rho}_{\mu\nu}\neq 0$) and quadratic ($\theta^{(2)\rho\tau}_{\mu\nu}\neq 0$) deformations of Minkowski space. The extension of particle models on noncommutative space-times to linearly and quadratically deformed Minkowski spaces is now studied.

\paragraph{Acknowledgements}
One of the authors (JL) would like to thank Prof. Ge Mo-Lin and Dr. Cheng-Ming Bai for their fantastic hospitality during the conference at Nankai Mathematical Institute in Tianjin. We also acknowledge the support of KBN grant 1P03B 01828 and the EPSRC.

\end{document}